\documentclass[preprints,article,accept,moreauthors,pdftex]{Definitions/mdpi}

\firstpage{1} 
\pubvolume{1}
\issuenum{1}
\articlenumber{0}
\pubyear{2021}
\copyrightyear{2020}
\datereceived{} 
\dateaccepted{} 
\datepublished{} 
\hreflink{https://doi.org/} 

\Title{Neutron-Mirror Neutron oscillations in Matter}

\TitleCitation{Neutron-Mirror Neutron oscillations in Matter}

\Author{Yuri Kamyshkov $^{1}$*, 
James Ternullo $^{1}$, 
Louis Varriano $^{2}$ and
Zurab Berezhiani $^{3,4}$}

\AuthorNames{Yuri Kamyshkov, James Ternullo, Louis Varriano and Zurab Berezhiani}

\AuthorCitation{Yuri Kamyshkov, James Ternullo, Louis Varriano and Zurab Berezhiani,}

\address{%
$^{1}$ \quad Department of Physics and Astronomy, The University of Tennessee, Knoxville, TN 37996, USA\\
$^{2}$ \quad Department of Physics, University of Chicago, Chicago, IL 60637, USA\\
$^{3}$ \quad Dipartimento di Fisica e Chimica, Università di L'Aquila, Coppito, 67100 L’Aquila, Italy\\
$^{4}$ \quad INFN, Laboratori Nazionali del Gran Sasso, 67010 Assergi 67100 L'Aquila, Italy}

\corres{Correspondence: kamyshkov@utk.edu}

\abstract{ The possibility that a neutron can be transformed to a hidden sector particle remains intriguingly open. Proposed theoretical models conjecture that the hidden sector can be represented by a mirror sector, and the neutron $n$ can oscillate into its sterile mirror twin $n'$, exactly or nearly degenerate in mass with $n$.
 Oscillations $n-n'$ can take place in vacuum and in the environment of the regular matter and the magnetic field where only neutron will be subject of interaction with the environment. We describe the propagation of the oscillating $n-n'$ system as a particle of the cold neutron beam passing through the dense absorbing materials in connection with the possible regeneration type of experiments where the effect of $n \rightarrow n' \rightarrow n$ transformation can be observed.}

\begin{document}
\section{Introduction}
\label{Sec:intro}

Current interest to the hidden sector in particle physics is motivated mostly by the apparent existence of Dark Matter (DM) 
which is not comprised in the Standard Model (SM) and 
the nature of which is yet not determined. One of the interesting possibilities is that DM is related to particles of some 
hidden sectors which also appear in string theory models. In these models hidden sector can include a new gauge group that is independent from the gauge group of the Standard Model or some of its 
extensions such as Grand Unification. Thus, the hidden sector particles do not interact with ordinary matter particles via Standard Model forces, however they have gravitational interactions common with ordinary matter. In the Mirror Matter model conjectured in \cite{Kobzarev:1966qya,Foot:1991bp}, 
hidden gauge sector is a replica of the ordinary sector including the same particles content so that two sectors are described by the 
Standard Model (SM) and its mirror copy SM$'$.  
They can have identical Lagrangians due to mirror $Z_2$ symmetry 
under the exchange of the particles between two sectors 
(for reviews, see 
\cite{Berezhiani:2003xm,Berezhiani:2005ek,Foot:2014mia}, 
and for a historical overview see \cite{Okun:2006eb}). 
Mirror matter can be a viable candidate for DM, 
with specific cosmological implications, 
provided that the temperature of the mirror sector is smaller 
than that of the ordinary sector \cite{Berezhiani:2000gw,Ignatiev:2003js,Berezhiani:2003wj}. 

If mirror symmetry $Z_2$ is an exact symmetry, 
i.e. the Higgs doublets of the SM and SM$'$ have exactly 
the same vacuum expectation values (VEV), 
$\langle \phi \rangle =\langle \phi' \rangle $, 
then the ordinary and mirror sector should have identical  
particle spectra, so that all ordinary particles: 
the electron $e$, proton $p$, neutron $n$ etc. have 
mass degenerate mirror twins $e'$, $p'$, $n'$ etc.  
which are sterile with respect to the SM interactions but have 
their own SM$'$ interactions. However, $Z_2$ symmetry can be 
spontaneously broken with two Higgses having different VEVs 
$\langle \phi \rangle \neq \langle \phi' \rangle $ in which case 
the mirror particle will have masses different from that of their 
ordinary partners \cite{Berezhiani:1995am,Berezhiani:1996sz}. 

Besides gravity, there can exist other interactions between the ordinary and mirror particles, possibly  giving rise to observable effects. 
The cross-interactions which violate the lepton and/or baryon numbers of both sectors are of particular interest. From one side, since these interactions violate both $B-L$ and $B'-L'$ symmetries, they can induce baryon asymmetries in both sectors \cite{Bento:2001rc,Bento:2002sj}, and such co-genesis mechanisms 
can explain the dark matter fraction in the Universe \cite{Berezhiani:2008zza,Berezhiani:2018zvs}.
On the other side, they can induce 
the mixing and oscillation phenomena between the neutral particles from both sectors, e.g. the neutrino mixing $\nu - \nu'$ between two sectors, which makes mirror neutrinos to be the natural candidates for sterile neutrinos \cite{Akhmedov:1992hh,Foot:1995pa,Berezhiani:1995yi}.
Also, the neutron $n$ can be mixed with a sterile neutron $n'$, 
its partner from the mirror sector, $\epsilon \bar{n} n' + {\rm h.c.}$ 
Interestingly, the oscillation $n-n'$ can be a rather fast process, with the characteristic oscillation time $\tau_{nn'}=\epsilon^{-1}$ as small as a few seconds: this possibility does not contradict the existing astrophysical limits, and it does not lead to the nuclear instability \cite{Berezhiani:2005hv},\footnote{In difference from  
the neutron--antineutron oscillation ~\cite{Kuzmin:1970nx,Mohapatra:1980de} for which the characteristic time should be $\tau_{n\bar n} > 10^8$~s as it is restricted by the direct experimental bound as well by the nuclear stability limits \cite{Phillips:2014fgb}.}.
This could also have interesting astrophysical implications e.g. for the extreme energy cosmic rays \cite{Berezhiani:2006je,Berezhiani:2011da} and for the neutron stars \cite{Goldman:2019dbq,Berezhiani:2020zck,Berezhiani:2021src}.

The reason why $n-n'$ transition faster than the neutron decay is not immediately manifested is that it is affected by the medium effects such as the presence of the matter or magnetic fields~\cite{Berezhiani:2005hv,Berezhiani:2008bc}. 
However, it can be observed via neutron disappearance $n\rightarrow n'$ or regeneration $n\rightarrow n' \rightarrow n$ \cite{Berezhiani:2005hv} in experiments with properly controlled background conditions. 
These experiments are convenient for observations due to the large lifetime of the neutron, the detection mechanism determined by strong interaction, and the large neutron fluxes available from the reactors or spallation sources.

In the previous works \cite{Berezhiani:2005hv,Pokotilovski:2006gq,Berezhiani:2017azg,Berezhiani:2018qqw} various possible experiments for observing $n-n'$ oscillation effects were considered, including essentially the two detection methods: the disappearance experiments due to the neutron oscillation $n\rightarrow n'$ into a sterile neutron $n'$, 
and the appearance (walking through the wall) experiments due to neutron regeneration $n\rightarrow n' \rightarrow n$ from the sterile state $n'$ with neutrons flowing through the absorbing wall.  

Several dedicated experiments were already performed to search for
$n \rightarrow n'$ oscillations via neutron disappearance in the ultra-cold neutron (UCN) traps \cite{Ban:2007tp,Serebrov:2008hw,Altarev:2009tg,Bodek:2009zz,Serebrov:2009zz,Berezhiani:2017jkn,nEDM:2020ekj}. These experiments still do not exclude the possibility of $n-n'$ oscillation time to be much less than the neutron decay time, and some of them even show anomalous deviations from null-hypothesis \cite{Berezhiani:2012rq}.  A new search is underway for testing these anomalies at the UCN facility of Paul Sherrer Insititute (PSI) \cite{nEDM2}. 

On the other hand, both the neutron disappearance $n\rightarrow n'$ and regeneration $n\rightarrow n' \rightarrow n$ can be experimentally tested with cold neutrons \cite{Berezhiani:2017azg}. 
The latter search can be realized e.g. in an experiment with an intense cold neutron beam where the transformation $n \rightarrow n'$ in the beam can be enhanced by applying specific environmental conditions. Then the neutron beam can be removed by a strong absorber leaving only $n'$ states passing freely through. After passing absorber, $n'$ can effectively oscillate back to the $n$ states under the same environmental conditions and will be detectable. Such experiments are underway at neutron sources in Oak Ridge National Laboratory (ORNL) \cite{Broussard:2017yev,Broussard:2019tgw, Broussard:2021} and at the newly constructed European Spallation Source (ESS) \cite{Addazi:2020nlz}. In the disappearance experiment a small reduction of the neutron flux should be detected under apply a certain constant magnetic field over the neutron flight distance. In the regeneration mode a small appearance effect can be directly measured possibly with a small background. For both methods, a resonant magnetic field should allow switching the effect ON/OFF.\footnote{Another type of experiment can be related to the neutron regeneration to the antineutron, $n\rightarrow (n',\bar{n}') \rightarrow \bar{n}$, which is possible if the neutron $n$ has mixings with both the mirror neutron $n$ and mirror antineutron $\bar{n}'$ \cite{Berezhiani:2020vbe}. }

The regeneration experiments are particularly promising for testing $n-n'$ oscillations 
in the case when the two states $n$ and $n'$ have some small mass splitting $\Delta m = m_n - m'_n \sim 100$ neV or so. 
In particular, this situation was used in \cite{Berezhiani:2018eds} 
for explaining the neutron lifetime discrepancy between the trap and beam experiments. Such a small splitting between the ordinary and mirror particles can be obtained if mirror symmetry $Z_2$ is very mildly broken, with the VEVs of two Higgses having a small difference, $\langle \phi' \rangle \approx \langle \phi \rangle $ ~\cite{Mohapatra:2017lqw,Berezhiani:2018udo}.

The model with mass splitting between $nn'$ and $n'$ is being tested with the cold neutrons in a strong magnetic field by the $NN'$ Collaboration at the Spallation Neutron Source in Oak Ridge National Laboratory by the regeneration method ~\cite{Broussard:2021}. An essential element of the regeneration method is the absorber, where a two-component oscillating $(n,n')$ system will propagate with only one component $n$ strongly interacting with the material environment and the other component, $n'$, is sterile. To our knowledge, no Quantum Mechanical consideration of the evolution of such a system, through strongly absorbing materials, have been reported. 
\footnote{The reflection/absorption of oscillating $(n,n')$ system in the trap with material walls was considered for UCN neutrons in paper \cite{Kerbikov:2008qs}.}. In view of the mentioned above experiments with cold neutrons, including regeneration search,  we performed this study.
\section{Description of Approach}
\label{Sec:approach}

 We shall describe the propagation of the oscillating $(n-n')$ system in the cold beam of neutrons, i.e. neutrons with a spectrum of velocities ranging practically from 200 m/s to 2000 m/s, through dense materials in view of the possible regeneration type experiments~\cite{Berezhiani:2017azg} where the effect of $n \rightarrow n' \rightarrow n$ transformation can be measured. The regeneration experiment can be described in the following way. The beam of free cold neutrons that can oscillate between $n$ and $n'$ states is propagating in a vacuum. Before entering the absorber it will have a probability $P_n$ to be detected as $n$ neutron (if neutron detector would be provided at this place) and the probability $P_{n'}=1-P_n$ to be in $n'$ state (since $n'$ is not detectable as such). The thickness of the absorber can be sufficiently large to remove all neutrons from the beam. The $n'$ component should pass through absorber without interaction. After exiting the absorber $n'$ can continue free oscillations, enriching the beam with $n$ component. The latter travelling through some distance to the detector will be counted there. Thus, neutrons can be found passing through the absorption wall. 
 
 Evolution of $n-n'$ system is described by Schr\"odinger equation 
$i d\Psi/dt = H \Psi$ where 
\begin{equation}\label{Psi-t}
\Psi(t)  = \begin{pmatrix}  \psi_n(t) \\ \psi_{n'}(t)  \end{pmatrix} 
\end{equation} 
where each of components $\psi_n=(\psi_n^+,\psi_n^-)^T$ and $\psi{n'}=(\psi_{n'}^+,\psi_{n'}^-)^T$ are in itself 2-component spinors describing the two spin states of the ordinary and mirror neutron. Hence, $\vert \psi_n(t) \vert^2 = \vert \psi_n(t)^+ \vert^2 + \vert \psi_n^-(t) \vert^2$ corresponds to the probability of
detecting the neutron at time $t$. 
A generic Hamiltonian in a medium has the form 
\begin{equation}\label{eq:24}
 H=\left(\begin{array}{cc}
E & \epsilon{} \\
\epsilon{} & E'
\end{array}\right)
\end{equation}
where $\epsilon$ is the $n-n'$ mixing mass, and $E$ and $E'$ are 
the energy levels for $n$ and $n'$ states corresponding to 
the same momentum $p$. 
Namely, in non-relativistic approximation one has 
\begin{equation} \label{eq:25}
E=m_n+ \frac{p^2}{2m} + {\mu{}}_n(\vec{\sigma{}}\, \vec{B}) +V -iW -i\frac{\Gamma_n}{2} 
\end{equation}
where $\mu_n$ is the neutron magnetic moment, $\vec{B}$ is the magnetic field and $\vec{\sigma}=(\sigma_1,\sigma_2,\sigma_3)$ are the Pauli matrices, $V$ is the neutron optical potential,
$W$ is the neutron absorption rate in matter, and 
$\Gamma_n=\tau_{\rm dec}^{-1}$ is the neutron decay rate.  
Similarly, $E'$ can be expressed in terms of a contribution from a mirror magnetic field and mirror matter material density, if the latter can be present in the experiment.

We assume that there is a mass splitting between ordinary and mirror neutrons, $\Delta{m} = m_{n'} - m_n$, which we take for the sake 
of definiteness be normalized to the values suggested in Ref. \cite{Berezhiani:2018eds} for solving the neutron lifetime problem, 
i.e. $\Delta m \sim 100$~neV or so. 
The identical real contributions in $E$ and $E'$ are 
irrelevant for the evolution of the system and we can omit them. 
We also assume that contributions of mirror matter and mirror magnetic fields are negligible and set $V',W'$ and $B'$ to zero. 
In this case, it is convenient to take the spin quantization axis as the direction of the magnetic field $\vec{B}=0,0,B)$. 
Therefore, the magnetic field contribution for two polarization states will be $+\mu_n B$ and $-\mu_n B$. 
In addition, the last term in eq.~(\ref{eq:25}) is the width of the neutron decay, which should be practically the same for neutrons and mirror neutrons. Since the neutron decay time ($\tau_{\rm dec}\approx 880$~s) is very large as compared to the cold neutron observation time that is typically $t\sim 0.1$ s, we can neglect this term in eq.~(\ref{eq:25}) and in the Hamiltonian.  
Therefore, our Hamiltonian (\ref{eq:24}) can be  
reduced to the following effective Hamiltonian
\begin{equation}\label{eq:18}
{\cal H}= \left(\begin{array}{cc}
U-iW & \epsilon \\
\epsilon  & 0
\end{array}\right) 
= \left(\begin{array}{cc}
-\Delta m \pm \mu_n B +V -iW & \epsilon{} \\
\epsilon{} & 0
\end{array}\right)
\end{equation}
which is non-hermitian in the presence of the absorptive contribution. Diagonal real quantities in eq.~(\ref{eq:18}) we combined into $U=-\Delta{m} \pm \mu_{n}B+V$.  This Hamiltonian can also be split in two parts, 
hermitian and anti-hermitian (absorptive):
\begin{equation}\label{HH}
{\cal H}= {\cal H}_{\rm osc} + {\cal H}_{\rm abs}, \quad 
{\cal H}_{\rm osc} = \left(\begin{array}{cc}
U & \epsilon \\
\epsilon & 0
\end{array}\right)  \quad  
{\cal H}_{\rm abs}= -i \left(\begin{array}{cc}
W & 0 \\
0 & 0
\end{array}\right)
\end{equation}

The Hamiltonian ${\cal H}$ is not hermitian if $W\neq 0$. 
But it can be diagonalized by a canonical transformation to 
$S^{-1} {\cal H} S = {\cal H}_{\rm diag}={\rm diag}(E_1,E_2)$, 
or vice versa, ${\cal H} =S {\cal H}_{\rm diag} S^{-1}$.  
The eigenvalues $E_{1,2}$ are generally complex. 
Without loss of generality, the matrix $S$ can be 
taken as unimodular:   
\begin{equation}\label{S} 
S= \begin{pmatrix}  c  & s  \\ -s  & c  \end{pmatrix},
\quad\quad c= \cos\zeta =\frac12\big(e^{i\zeta} + e^{-i\zeta}\big), \quad 
s= \sin \zeta = \frac{1}{2i}\big(e^{i\zeta} - e^{-i\zeta}\big) 
\end{equation}
with the parameter $\zeta$ being generally complex. 
(Once again, if the absorptive part $W$ is vanishing, then $\zeta$ is real and the matrix $S$ becomes unitary). 

Then the Schr\"odinger equation $i d\Psi/dt = {\cal H}\, \Psi$ in a constant environment 
is formally solved as 
\begin{equation}\label{Hexp}  
\Psi(t) = {\mathcal S}(t) \Psi(0) , \quad
{\cal S}(t) =  e^{-i {\cal H} t } =
 e^{-i S {\cal H}_{\rm diag} S^{-1} t}  
= S \,{\rm diag} (e^{-i E_1 t}, e^{-i E_2 t})\, S^{-1}
\end{equation}
where the overall phase factor development 
is described solely by the evolution matrix 
\begin{equation}\label{S-matrix} 
 {\cal S}(t) 
 = \begin{pmatrix}  {\cal S}_{nn}(t)  & {\cal S}_{nn'}(t) \\ 
 {\cal S}_{n'n}(t) & {\cal S}_{n'n'}(t) \end{pmatrix} 
 = \begin{pmatrix}  c^2 e^{-iE_1t} + s^2 e^{-iE_2t} & cs (e^{-iE_2t} - e^{-iE_1t}) \\ 
 cs (e^{-iE_2t} - e^{-iE_1t})  & s^2 e^{-iE_1t} + c^2 e^{-iE_2t}
 \end{pmatrix} 
\end{equation} 

In this paper we are following the approach of our previous paper 
\cite{Berezhiani:2018qqw}. Also, we noticed that interaction of oscillating 
($n,\bar{n}$) system with magnetic and gas environment was 
considered in the paper \cite{Gudkov:2019gro}.
For our purposes it will be convenient to use the density matrix 
formalism and describe the evolution of 
 $(n,n')$ system evolution via the Liouville–von Neumann equation
 (see for example \cite{Breuer:2002}):
\begin{equation}\label{eq:21}
\dot{\rho{}}=-i\big({\cal H} \rho -\rho {\cal H}^{\dagger} \big) 
= - i \big[{\cal H}_{\rm osc}, \,\rho\big] 
-i \big[{\cal H}_{\rm abs}, \,\rho\big]_+
\end{equation}
where $\rho(t)$ is a $4 \times 4$ Hermitian density matrix:  $\rho_{i,j}(t)=\Psi_{i}(t) \Psi^{*}_{j}(t)$, where $i,j = n, n'$. Diagonal terms of density matrix $\rho_{nn}(t)$ and $\rho_{n'n'}(t)$ represent the probability of observation at time $t$ of neutron and sterile neutron correspondingly.
The first term in~eq.(\ref{eq:21}) is related to the Hermitian part of the Hamiltonian contains a commutator. The second term related to its absorptive part contains an anti-commutator. We set Tr$(\rho)=1$ at initial $t=0$.

We should notice that the Hamiltonian eq.~(\ref{eq:24}) describing te interaction of $(n,n')$ system with an absorber is incomplete. It doesn't include effects of scattering of neutrons on the nuclei of the absorber material. Since mirror neutrons do not scatter off nuclei, the scattering at any angle different than zero will lead to decoherence of the oscillating $(n,n')$ system. This decoherence should be properly treated with more complicated Lindblad Master equation \cite{Lindblad:1975ef,Gorini:1975nb}. Similar treatment for oscillating muonium-antimuonium system in a gas environment was considered in an early paper \cite{Feinberg:1961zza}. For the regeneration effect that we are considering in this paper, the elastic scattering of neutrons at angles larger than zero will additionally reduce the number of neutrons at coordinate z that are removed due to absorption. For the mirror neutron component entering the absorber, transitional oscillations are being damped at the neutron absorption length which is much smaller than the length for elastic scattering. Thus, if e.g. cadmium is used as an absorber, the absorption length is $\sim 0.04 mm$ while the elastic scattering length is $\sim 20 mm$.

In the next section we will start with discussion of the preparation of the initial state of the density matrix in a vacuum and calculation of average density matrix as initial state. Next, we will discuss the case when a magnetic field is absent, i.e. $U=-\Delta{m}+V_{op}$ and will calculate, numerically, the evolution of the density matrix for an idealized regeneration experiment where a neutron is passing the vacuum-absorber-vacuum-detector environment sequence. If the regeneration experiment will occur in a constant magnetic field, then this can be described in the same way as a zero magnetic field but with modification of the magnitude of $U$, and by considering two possible beam polarizations with $-\mu{B}$ and $+\mu{B}$. Finally, we will discuss the calculations with constant $U$ for the case of weak and strong absorbers. The case with $U(z)$ including $\Delta{m}$ and complex $V_{opt}$ in non-uniform magnetic field $B(z)$ was used in calculations for the regeneration experiment \cite{Broussard:2021} and not presented in this paper.

\section{Average Density Matrix in Vacuum}
\label{Sec:Average}
Now, following the paper~\cite{Berezhiani:2018eds}, we can consider an oscillating $(n,n')$ system with $U=-\Delta{m}$ propagating in a vacuum, with $B=0$, $V=0$, and $W=0$. 
\begin{equation}
{\cal H} =\left(\begin{array}{
cc}
-\Delta{m} & \epsilon{} \\
\epsilon{} & 0
\end{array}\right)\label{eq:27}
\end{equation}

This Hamiltonian will be used in eq.~(\ref{eq:21}) together with the density matrix 

\begin{equation}
\rho(t)=\left(\begin{array}{
cc}
\rho_{nn}(t) & \rho_{nn'}(t) \\
\rho_{n'n}(t) & \rho_{n'n'}(t)
\end{array}\right)\label{eq:28}
\end{equation}

At the point where the neutron is produced in the nuclear reactor or in spallation process, or as a result of neutron scattering or decay of other particles, the initial state of the density matrix for $(n,n')$ system at $t=0$ can be described as 
\begin{equation}
\rho(0)=\left(\begin{array}{
cc}
1 & 0 \\
0 & 0
\end{array}\right)\label{eq:29}
\end{equation}

Time evolution of the density matrix eq.(\ref{eq:28}) will depend on the oscillation frequency $\omega$ and the mixing angle $\theta_{0}$ defining the amplitude of oscillations
\begin{equation}
\omega=\sqrt{(\Delta{m}/2)^2+\epsilon^2}
\label{eq:30}
\end{equation}
\begin{equation}
\mathrm{tan}2\theta_{0}=-2\epsilon/\Delta{m}
\label{eq:31}
\end{equation}
The solution of eq.(\ref{eq:21}) under initial conditions of eq.~(\ref{eq:29}) can be explicitly found by
\begin{equation}
\begin{split}
{\rho{}}_{nn}(t)=1-{sin}^2 2{\theta{}}_0\cdot{sin}^2 \omega{}t \\
{\rho{}}_{nn'}(t)=-\frac{1}{2}sin{4\theta}_0\cdot{sin}^2{}\omega{}t-\frac{i}{2}sin2{\theta}_0\cdot sin2\omega{}t \\
{\rho{}}_{n'n}(t)={\rho{}}_{nn'}^*(t) \\
{\rho{}}_{n'n'}(t)={sin}^2 2{\theta{}}_0\cdot{sin}^2 \omega{}t
\label{eq:32}
\end{split}
\end{equation}

If $\Delta{m}>10~neV$ then an for arbitrarily small $\epsilon$ the frequency $\omega$ will be large enough such that for the neutron moving in the Lab with velocity $\sim{~1000~m/s}$ will have an oscillation length smaller than 1 mm. If the size of the experiment is much larger than 1 mm then the time dependent probabilities of eq.~(\ref{eq:32}) can be replaced by time-averaged values. Thus, we are coming to the time-averaged density matrix (TADM). In the TADM, the oscillation phases induced by variable initial phases and velocities of the beam neutrons, will be averaged in the following way (we also using here the smallness of the angle $\theta_{0}$)
\begin{equation}
\bar{\rho}=\left(\begin{array}{
cc}
1-2{\theta}^2_0 & -\theta_0 \\
-\theta_0 & 2{\theta}^2_0
\end{array}\right)\label{eq:33}
\end{equation}
This matrix when taken as the initial condition for equation eq.~(\ref{eq:21}) with the Hamiltonian eq.~(\ref{eq:27}) reproduces itself in evolution. 
This density matrix eq.~(\ref{eq:33}) is providing a phase averaging and can be used as an initial condition for evolution of the $(n,n')$ system coming from vacuum through the absorbing material. This density matrix can be understood as a final state of a single $(n,n')$ system having probabilities similar to that of the average of the large ensemble of neutrons in the beam.
 \section{Evolution through weakly absorbing material}
 \label{Sec:weakly}
As a practice of application of cold neutrons, sometimes it is necessary to transport the neutron beam through air. This is an example of a weakly absorbing environment for propagation of the $(n,n')$ system.
We will try to construct an average density matrix that will be a result of evolution through homogeneous weakly-absorbing material described by the Hamiltonian eq.(\ref{eq:18}) without a magnetic field, such that $W=0$ and $U=-\Delta{m}+V_{opt}$.

For propagation of cold neutrons in air, at NTP we used the following values for  $V_{opt}=5.668\times10^{-11}$ eV and $W_{opt}=(5.773\times10^{-15}+v\cdot8.328\times10^{-18}$) eV, where $v$ is neutron velocity in m/s. By direct calculation of $\rho(t)$ in evolution of eq.(\ref{eq:21}) 
we obtained the following density matrix for the $(n,n')$ system passing path in air  $\Delta{z}=v\cdot\Delta{t}$. As we mentioned before, we are not considering in the evolution the elastic scattering of cold neutrons off the nuclei in the gas, since elastic scattering will result in a dropout of neutrons from a highly-collimated beam and effectively reduce the beam intensity, but will not fundamentaly affect the propagation of $(n,n')$ system.

\begin{equation}
\begin{split}
{\bar{\rho{}}}_{nn}=\left(1-\tfrac{1}{2}\cdot{} sin^2 2{\theta_0}\right)\cdot{}e^{-\Delta{z}/L_{air}} \\
Re{\bar{\rho{}}}_{nn'}=Re{\bar{\rho{}}}_{n'n}=\tfrac{1}{4}\cdot{}sin4\theta_{0}\cdot{}e^{-\Delta{z}/L_{air}} \\
Im{\bar{\rho{}}}_{nn'}=Im{\bar{\rho{}}}_{n'n}=0 \\ 
{\bar{\rho{}}}_{n'n'}=\tfrac{1}{2}\cdot{} sin^2 2{\theta_0}\cdot{}e^{-2\Delta{z}/L_{air}} \\
\label{eq:34}
\end{split}
\end{equation}
,where $L_{air}$ is absorption length in air calculated as $L_{air}=\hbar{}v/(2W)$. For example, for neutron velocity $v=1000$ m/s absorption length $L_{air}=23.338$ m.
The averaged density matrix in eq.(\ref{eq:34}) can be used as the initial state of the evolution for the $(n,n')$ system entering a strongly-absorbing material after passing the distance $\Delta{z}$ in air or in another weakly absorbing media.
\section{Evolution in the strongly absorbing material}
\label{Sec:strongly}
\begin{figure}[H]
  \begin{center}
  \includegraphics[width=13.0 cm]{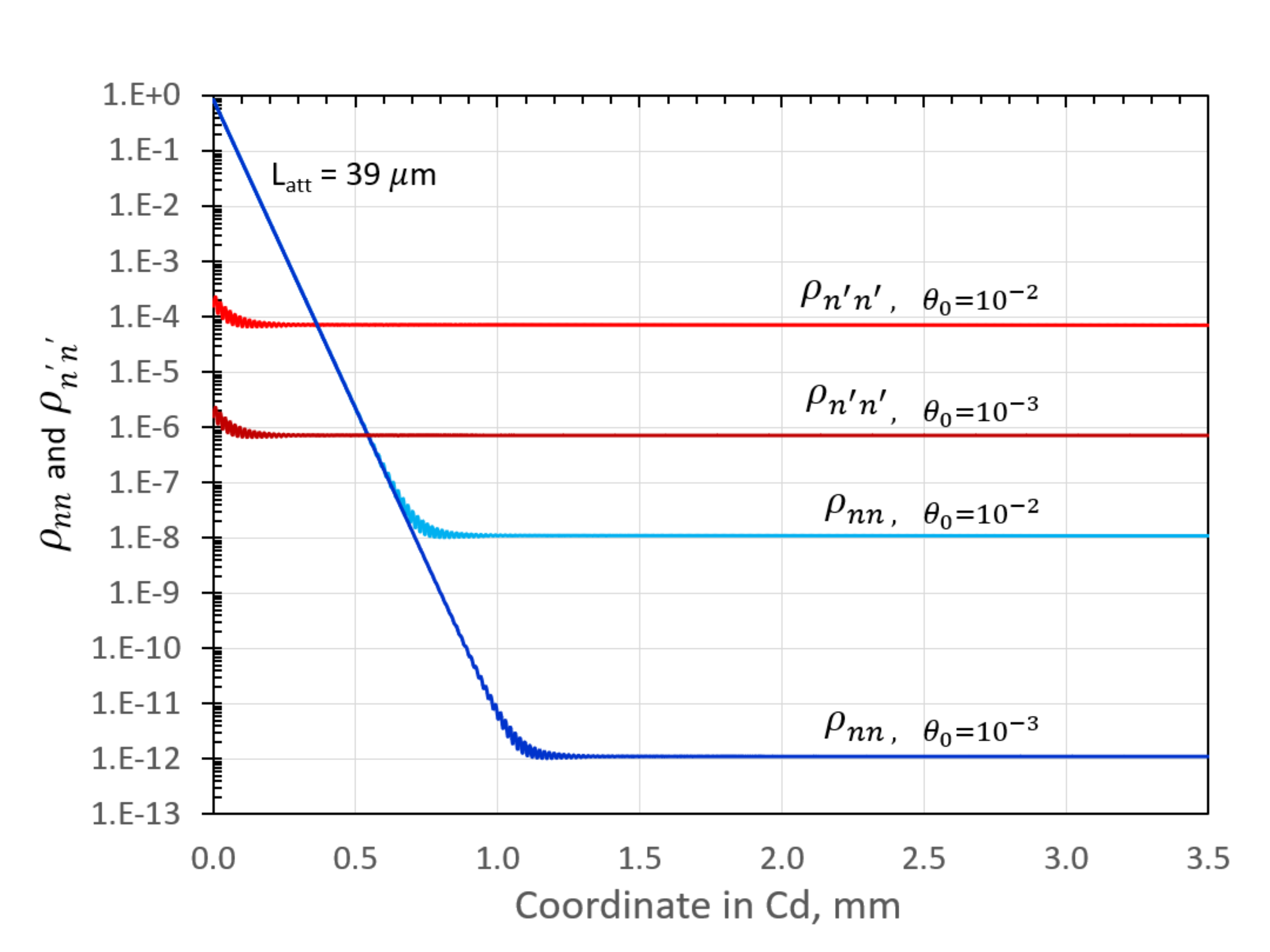}
  \end{center}
  \caption{Evolution of $(n,n')$ system in 3.5 mm Cd absorber for $\Delta{m}=300$ neV, ${\theta}_{0}=10^{-2}$ and $10^{-3}$, and for velocity $v=1000$ m/s. Averaged density matrix at the entrance of Cd-absorber after passing 3-m in air is used as an initial condition. Magnetic field $B=0$. Neutron components ${\rho}_{nn}$ are shown in light/dark blue and mirror neutron component ${\rho}_{n'n'}$ in light/dark red colors.}
  \label{fig:1}
\end{figure}

Evolution of the $(n,n')$ system inside the absorber can be described by the same Hamiltonian eq.(\ref{eq:18}) without a magnetic field with different values for $V_{opt}$ and $W_{opt}$ for the particular absorbing material. As example calculation for practical reasons, we have chosen two absorber materials: a 3.5 mm thick Cadmium ($Cd$) and 32 mm Boron Carbide ($B_{4}C$) with natural isotope abundance. For Cd $V_{opt}=5.877\times{}10^{-8} eV$ and $W_{opt}=(8.4558\times{}10^{-9}+v\cdot{}9.914\times{}10^{-15})eV$. For $B_{4}C$ $V_{opt}=1.992\times{}10^{-7} eV$ and $W_{opt}=(6.102\times{}10^{-9}+v\cdot{}2.397\times{}10^{-14}) eV$. An example of evolution calculations for Cd is shown in fig.\ref{fig:1} for parameters $\Delta{m}=300~neV$ and $\theta_{0}=1\times{}10^{-2}$, and $\theta_{0}=1\times{}10^{-3}$.

As expected, the mirror neutron component ($\rho{}_{n'n'}$) remains practically constant throughout the depth of the absorber, but at the entrance it experiences some damped oscillations due to re-arrangement of the energy eigenvalues of the system. The neutron component ($\rho{}_{nn}$) shows fast absorption with conventional absorption length known from the neutron cross sections until it reaches the level that is determined by oscillation feedback of mirror neutrons to neutrons. Since the probability of $n'$ remains almost constant it provides an equilibrium level of neutron probability. Oscillation probability is slightly damped in this equilibrium, however very small, invisible in the figure. In equilibrium the attenuation of both components is near equal. What is interesting is that the probability for $n'$ levels is approximately $\theta{}^2_{0}$, while the constant level of probability for neutrons is $\sim{}\theta{}^4_{0}$, thus providing regeneration effect already inside the absorber. Since oscillations are suppressed inside the absorber the system does not obtain the phase shift factor that would lead to the significant variation in the probability at the exit of the absorber, thus providing the state of the density matrix that is averaged, and can be served as an initial condition for the propagation in a vacuum (or in air, or in magnetic field) behind the absorber. 

\begin{figure}[h]
  \begin{center}
  \includegraphics[width=13.0 cm]{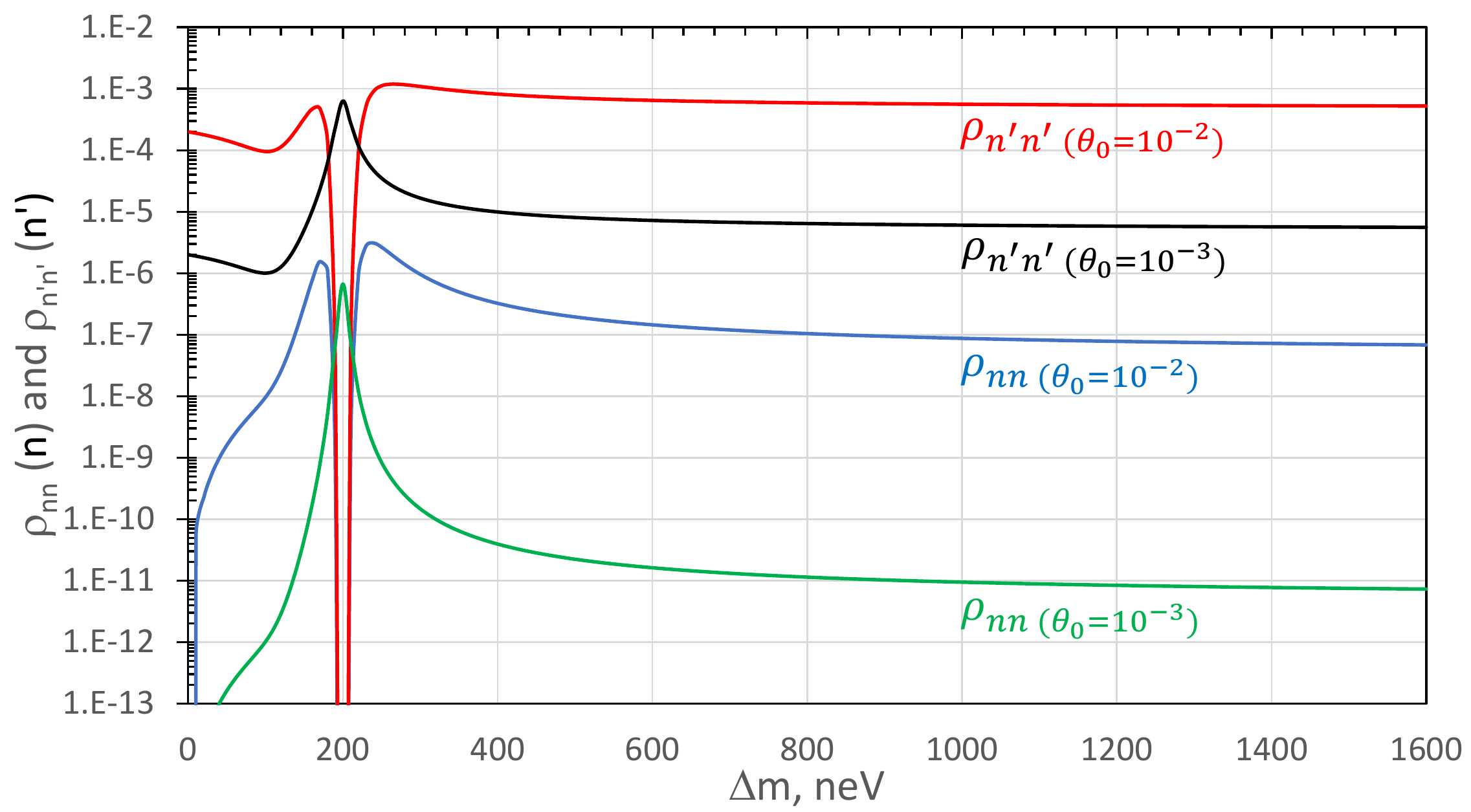}
  \end{center}
  \caption{$\Delta{m}$ dependence of probability $\rho_{nn}$ and $\rho_{n'n'}$ for two values of angles $\theta_0=0.01$ and $\theta_0=0.001$ after the neutron passes the 32 mm $B_{4}C$ absorber starting from the average density matrix in a vacuum. The value for the magnetic field $B=0$.}
  \label{fig:2}
\end{figure}

This interesting behaviour of the $(n,n')$ system inside the absorber opens up a new, very simple way of observation for the presence of mirror neutrons in the intense beam of cold neutrons. For that, it should be sufficient to measure the attenuation of the neutron beam intensity as a function of the absorber thickness. At some thickness the attenuation regime should be stopped and replaced by a constant irreducible intensity. The latter should be above the level of the background in the neutron detector. Since the constant level of neutron probability in essentially determined by the $\sim{}\theta{}^4_{0}$, such measurements will be not difficult to perform for larger values of $\theta{}_{0}$. The parameter $\theta{}_{0}=\epsilon{}/\Delta{m}$ can be limited by measurements for all values of $\Delta{m}$.
\section{Oscillation resonance in the strongly absorbing material}
\label{Sec:resonance}
\begin{figure} [H]
  \begin{center}
  \includegraphics[width=13.0 cm]{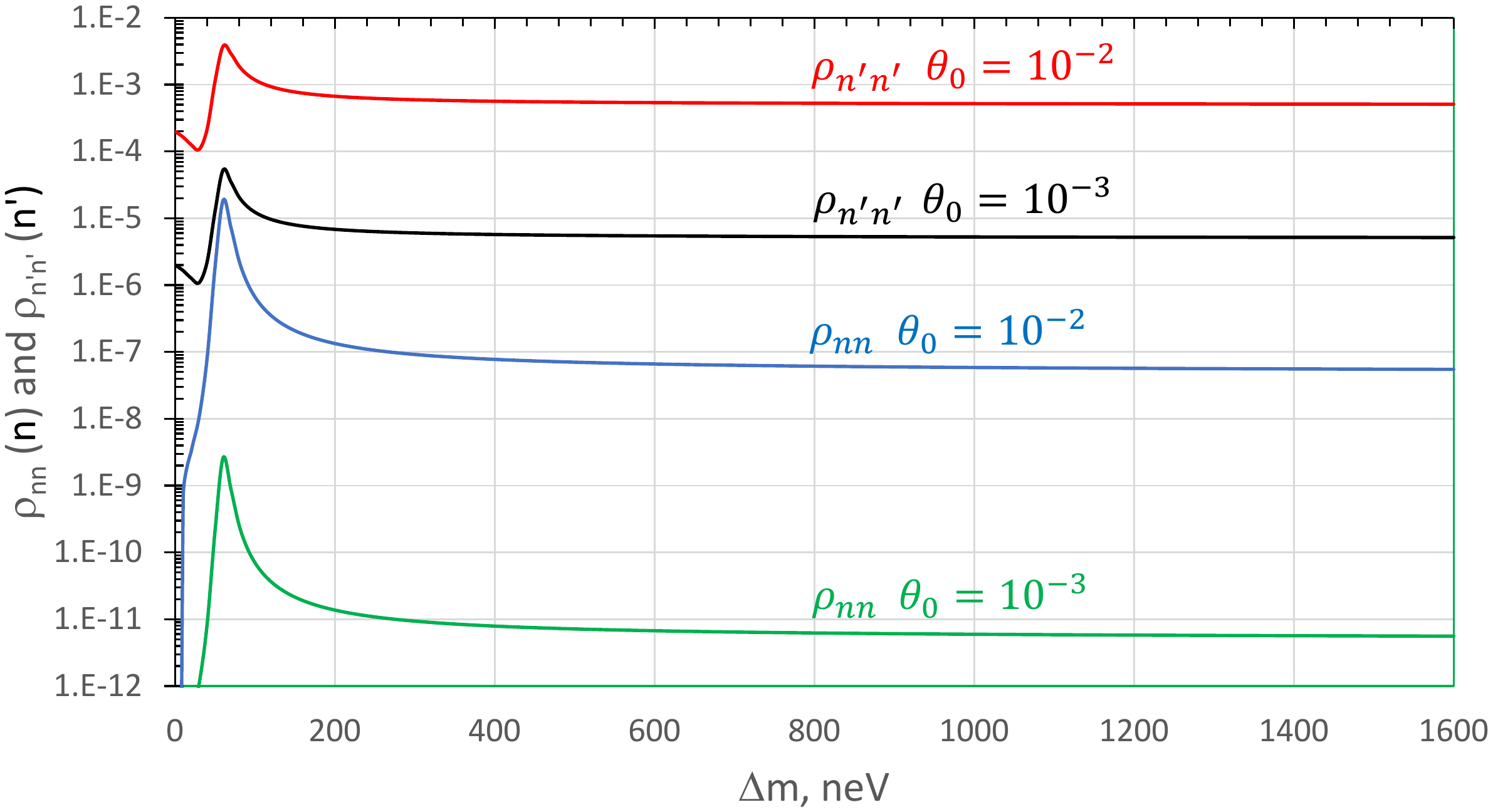}
  \end{center}
 \caption{$\Delta{m}$ dependence of probability $\rho_{nn}$ and $\rho_{n'n'}$ for two values of angles $\theta_0=0.01$ and $\theta_0=0.001$ after neutron passing 3.5 mm in $Cd$ absorber starting from averaged density matrix in vacuum. Magnetic field $B=0$.}  
  \label{fig:3}
\end{figure}

One can notice that in eq.~(\ref{eq:18}), without a magnetic field, $U=-\Delta{m}+V_{opt}$ and there is a possibility that $U$ can become $=0$ ~and the oscillation frequency modified to
\begin{equation}
\omega=\sqrt{(\theta{}_{0}\Delta{m})^2-(W/2)^2}
\label{eq:49}
\end{equation}
 This case also can be calculated with the evolution equation ~(\ref{eq:21}) and with the Hamiltonian in eq.~(\ref{eq:18}). We show the results of such calculations as the values of $\rho{}_{nn}$ and $\rho{}_{n'n'}$ in fig. \ref{fig:2} for a $B_{4}C$ 32-mm absorber and in fig. \ref{fig:3} for a $Cd$ 3.5-mm absorber. For both figures we calculated the density matrix components at the exit of the absorber as a function of $\Delta{m}$ for two values of $\theta{}_{0}=$~0.01 and 0.001.

The Fermi potential of $B_{4}C$ is $V_{opt}(B_{4}C)=1.992\times{}10^{-7} eV$ corresponding to a pronounced structure of probability in fig. \ref{fig:2} around $\Delta{m}=200 neV$. For a small angle $\theta{}_{0}$,
resonance enhances the regeneration of neutrons $\rho_{nn}$ and increases the yield of mirror neutrons $\rho_{n'n'}$. For larger angles at a stronger mixing parameter the absorption of neutron components starts to dominate in the resonance region. 

As we mentioned above, the presence of a constant magnetic field can play the same role in $U$ as $\Delta{m}$ or the real part of the optical potential $V_{op}$. Therefore, the position of the resonance can be controlled for some region of $\Delta{m}$ and the magnetic field. For $Cd$ absorber where $V_{op}=58.8 neV$ similar resonance behavior fig. \ref{fig:3} is not as pronounced as with $B_{4}C$ for the same values of mixing angle $\theta_{0}$.

\section{Summary}
By computing the time evolution of the density matrix of the two-level system $(n,n')$ passing the environment where one of the components in the Hamiltonian is strongly interacting with an environment and another component is sterile, we provided the method for understanding the process of regeneration that can be used in experiments with cold neutron beams, e.g. in~\cite{Broussard:2021}.
The real part of the potential of the Hamiltonian describing the two-level $(n,n')$ oscillating system interacting with the environment can include the effect of $\Delta m_{nn'}$ that can be positive and/or negative, the positive optical potential of the material, and the magnetic field whose contribution will depend on the polarization of the neutron. If a magnetic field varies along the path of the neutron beam 
it may compensate the overall real potential to zero and will lead to the resonance behavior inside the absorber that might essentially modify the regeneration process. This method can also be applied to the experiments with lower magnetic fields where $n$ and $n'$ are degenerate in mass but the mirror magnetic field $B'$ and/or eventual mirror matter gas can contribute to the energy splitting. In this case the latter contribution should not be neglected in the Hamiltonian \eqref{eq:24}.

\section{Acknowledgments}

The work of Z.B. was supported in part by the research grant 
``The Dark Universe: A Synergic Multimessenger Approach" No. 2017X7X85K under the program PRIN 2017 funded by the Ministero dell'Istruzione, Universit\`a e della Ricerca (MIUR),  
and  in part by Shota Rustaveli National Science Foundation 
(SRNSF) of Georgia, grant DI-18-335/New Theoretical Models for Dark Matter Exploration. L.V. was supported by a National Science Foundation Graduate Research Fellowship under Grant No. DGE-1746045.

\end{paracol}
\newpage
\section{References}
\label{Sec:References}
\bibliography{nnbar.bib} 
\end{document}